\documentclass{article}
\usepackage[margin=2cm]{geometry}
\usepackage{amssymb}
\usepackage{natbib}
\begin{document}

\title{Open problem:\\Computing Bayes-Nash equilibrium in simple
  security games\thanks{Originally prepared for the program
    ``Economics and Computation'' held in the Fall 2015 semester at
    the Simons Institute for the Theory of Computing.}}
\author{Vincent Conitzer\\ \texttt{conitzer@cs.cmu.edu} } \date{}
\maketitle

\section{The problem}

A {\em Bayesian simple security game} is given by (all explicitly enumerated):
\begin{itemize}
\item a set of targets $T$
\item a number of resources $r$ ($r \leq |T|$) for the defender
\item a set of attacker types $\Theta$ (with a probability distribution over
  them)
\item for every target $t$, utilities $u_1^0(t)$ and $u_1^1(t)$ ($u_1^0(t)
  \leq u_1^1(t)$), and for every type $\theta$, utilities $u_2^0(\theta,t)$
  and $u_2^1(\theta,t)$ ($u_2^0(\theta,t) \geq u_2^1(\theta,t)$)
\end{itemize}
In the game the defender chooses a subset of $r$ targets to assign her
resources to, and the attacker (after learning his type $\theta$) chooses a
target $t$ to attack.  If no defensive resource was assigned to $t$, then
the defender and attacker's utilities are $u_1^0(t)$ and $u_2^0(\theta,t)$,
respectively.  If a defensive resource was assigned to $t$, then the
defender and attacker's
utilities are $u_1^1(t)$ and $u_2^1(\theta,t)$, respectively.\\

\noindent
{\bf Problem:} Can a (Bayes-)Nash equilibrium be computed in polynomial
time for Bayesian simple security games?

\section{Further background}

Security games were introduced by~\citet{Kiekintveld09:Computing} in
their work on applying these techniques to real security domains.
Usually, in these games, Stackelberg mixed strategies are computed
rather than Nash equilibria.  In two-player normal-form games
computing the former can be done in polynomial
time~\citep{Conitzer06:Computing,Stengel10:Leadership}, while Nash is
of course
PPAD-complete~\citep{Daskalakis09:Complexity,Chen09:Settling}.  But in
this context (where the normal form has exponential size but there is
structure) the situation seems reversed.  We (with Yuqian Li, Dima
Korzhyk, and Ron Parr) have an NP-hardness result for Stackelberg
mixed strategies, as well as a fairly complicated, seemingly fast
algorithm for (Bayes-)Nash equilibrium for which we can't prove
whether or not it always runs in polynomial time.  For a simpler,
non-Bayesian variant where the attacker has multiple resources, a
similar (somewhat simpler) algorithm does solve the problem in
polynomial time~\citep{Korzhyk11:Security}.  It also appears that in
generic games of this type, the (Bayes-)Nash equilibria of these games
satisfy the interchangeability property, so that equilibrium selection
is not an issue~\citep{ChatGPT26:Interchangeability}.

\end{document}